\documentclass[runningheads]{llncs}
\usepackage{graphicx}
\usepackage{amsmath,amssymb} %
\usepackage{mathrsfs}
\usepackage{color}
\usepackage{multirow}
\usepackage[misc]{ifsym}
\usepackage[pagebackref=true,breaklinks=true,letterpaper=true,colorlinks=true,bookmarks=false]{hyperref}
\usepackage[accsupp]{axessibility}

\makeatletter
\def\thanks#1{\protected@xdef\@thanks{\@thanks
        \protect\footnotetext{#1}}}
\makeatother

\begin{document}
\pagestyle{headings}
\mainmatter

\title{Modular Degradation Simulation and Restoration for Under-Display Camera}
\author{
Yang Zhou$^\dagger$  \thanks{$^\dagger$ Equal contribution} \and
Yuda Song$^\dagger$  \and
Xin Du {\textsuperscript \Letter}
}
\institute{Zhejiang University, Hangzhou, China\\
\email{\{yang\_zhou,syd,duxin\}@zju.edu.cn}}

\titlerunning{MPGNet for UDC Simulation}
\authorrunning{Zhou et al.}

\maketitle

\begin{abstract}
Under-display camera (UDC) provides an elegant solution for full-screen smartphones.
However, UDC captured images suffer from severe degradation since sensors lie under the display. 
Although this issue can be tackled by image restoration networks, these networks require large-scale image pairs for training.  
To this end, we propose a modular network dubbed MPGNet trained using the generative adversarial network (GAN) framework for simulating UDC imaging. 
Specifically, we note that the UDC imaging degradation process contains brightness attenuation, blurring, and noise corruption. 
Thus we model each degradation with a characteristic-related modular network, and all modular networks are cascaded to form the generator. 
Together with a pixel-wise discriminator and supervised loss, we can train the generator to simulate the UDC imaging degradation process.
Furthermore, we present a Transformer-style network named DWFormer for UDC image restoration.
For practical purposes, we use depth-wise convolution instead of the multi-head self-attention to aggregate local spatial information.
Moreover, we propose a novel channel attention module to aggregate global information, which is critical for brightness recovery.
We conduct evaluations on the UDC benchmark, and our method surpasses the previous state-of-the-art models by 1.23 dB on the P-OLED track and 0.71 dB on the T-OLED track, respectively.
Code is available at \href{https://github.com/SummerParadise-0922/MPGNet}{Github}.
\end{abstract}
\section{Introduction}
Driven by the strong demand for full-screen mobile phones, the under-display camera (UDC) increasingly draws researchers' attention. 
UDC technology can deliver a higher screen-to-body ratio without disrupting the screen's integrity and introducing additional mechanics. 
However, UDC provides a better user experience at the expense of image quality. Since the sensor is mounted behind the display, the UDC images inevitably suffer severe degradation. 
Such image degradation is mainly caused by low light transmittance, undesirable light diffraction, and high-level noise, resulting in dark, blurred, and noisy images. 

Following the prior work~\cite{zhou2021image}, the UDC imaging degradation process can be formulated as:
\begin{equation}
    y = (\gamma \cdot x)\otimes k + n,
\end{equation}
where $\cdot$ and $\otimes$ are multiplication and convolution operations, respectively, $\gamma$ is the luminance scaling factor under the current gain setting and display type, $k$ donates the point spread function (PSF)~\cite{heath2018scientific}, and $n$ is the zero-mean signal-dependent noise. 
Considering the attenuation of light transmission is wavelength-related~\cite{nayar1999vision}, the luminance scaling factor $\gamma$ should be different for each channel. 
Also, the PSF is spatially varying due to the different angles of the incident light~\cite{kwon2021controllable}. 
And the signal-dependent noise $n$ consists of shot and read noise~\cite{hasinoff2014photon} which can be modeled by heteroscedastic Gaussian distribution~\cite{kersting2007most}.

In recent years, many learning-based methods~\cite{panikkasseril2020transform,sundar2020deep,nie2020dual,feng2021removing} have been introduced to improve the quality of UDC images and made significant advancements as they can learn strong priors from large-scale datasets.
However, such methods require large amounts of data, and collecting aligned image pairs is labor-intensive. 
Facing insufficient data, generating realistic datasets can be a promising solution.

\begin{figure}[t]
\centering
\includegraphics[width=1.0\textwidth]{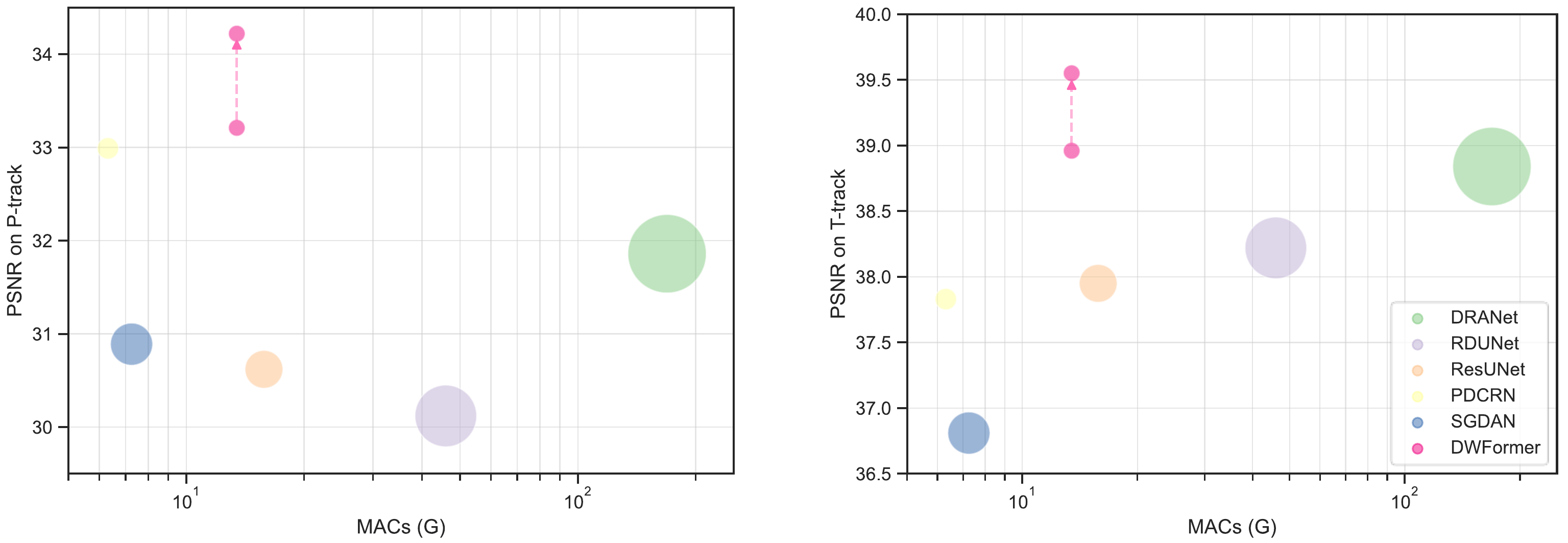}
\caption{
Comparison of DWFormer with other UDC image restoration methods. 
The size of the dots indicates the Params of the method, and MACs are shown with the logarithmic axis. 
The arrows indicate that we use the generated data to improve the restoration model's performance further.
}
\label{Fig.1}
\end{figure}

This paper proposes a modular pipeline generative network named MPGNet to simulate the UDC imaging degradation process and then use it to generate realistic image pairs.
Unlike other end-to-end generation methods~\cite{chen2018image,chang2020learning,kim2019grdn}, we replace each degradation process with a subnetwork to preserve the physical properties of the imaging process. 
Specifically, we treat the image degradation process as three sequential steps, \emph{i.e.,} brightness attenuation, non-uniform blurring, and noise corruption.
And all modular subnetworks form the UDC imaging pipeline network. 
Besides the supervised learning, we employ GAN framework~\cite{goodfellow2014generative} to enhance the realism of generated images.

Based on the large amount of data generated by MPGNet, we can obtain a restoration model with well-generalized performance.
However, designing a network suitable for the UDC image restoration task is not trivial. 
Recently, Transformers have attracted researchers' interest in the computer vision since ViT's~\cite{dosovitskiy2020image} success on the high-level vision tasks. 
Thus we hope to build an efficient and effective UDC image restoration network based on the vision Transformer.
Considering that MetaFormer~\cite{yu2021metaformer} reveals the general architecture of the transformers is more critical than the multi-head self-attention, we use depth-wise convolution and channel attention module to aggregate global information. 
Finally, we build a U-Net-like restoration network dubbed DWFormer for UDC image restoration.

We conduct evaluations on the UDC benchmark to verify the effectiveness of our MPGNet and DWFormer.
Fig.~\ref{Fig.1} compares DWFormer with other UDC image restoration methods.
Without synthetic datasets generated by MPGNet, DWFormer still achieves 33.21 dB on the P-OLED track and 38.96 dB on the T-OLED track, which surpasses the previous state-of-the-art models by 0.22 dB and 0.12 dB, respectively.
Furthermore, if we use both generated and real data to train our DWFormer, it achieves 34.22 dB on the P-OLED track and 39.55 dB on the T-OLED track, 1.01 dB and 0.59 dB higher than the DWFormer trained with only real data. 
These results indicate that MPGNet can precisely model the UDC imaging process, and DWFormer can restore UDC images effectively. 
We hope our work can promote the application of UDC on full-screen phones.

\section{Related Works}
\subsection{UDC Imaging}
Several previous works~\cite{zhou2021image,kwon2021controllable,2016Modeling,Zong2016See} have constructed the optical system of under-display camera (UDC) imaging and analyzed its degradation components as well as the causes. 
While these works provided good insights into the UDC imaging system, their modeling approaches simplify the actual degradation process. Thus the derived degradation images differ significantly from the real ones. 
Therefore, how to generate realistic degradation images is still a problem to be solved, and this is one of the focuses of our work. 
We found that some work has been done to study the degradation module individually.

\textbf{Blur Modeling.}
The blurring process can be modeled as a blurring kernel performing a convolution operation on a sharp image~\cite{whyte2012non,gupta2010single}. 
Many methods~\cite{sun2015learning,xu2014deep,chakrabarti2016neural} estimate the blur kernel by assuming the characteristics of the blur kernel, and other methods~\cite{zhou2021image,kwon2021controllable} models blur by a point spread function (PSF).
However, the blur in UDC imaging is blind and spatially varying, increasing the difficulty of accurately estimating the blur kernel. 
Unlike previous works, we try to model the blur directly using a convolutional neural network.

\textbf{Noise Modeling.}
The noise is usually modeled as Poissonian-Gaussian noise~\cite{foi2008practical} or heteroscedastic Gaussian~\cite{hasinoff2010noise}. 
While the heteroscedastic Gaussian noise model can provide a proper approximation of the realistic noise~\cite{guo2019toward} to some extent, several studies~\cite{wei2020physics,zhang2021rethinking} have demonstrated that the real-world cases appear to be much more complicated.
To this end, GCBD~\cite{chen2018image} proposed a generation-based method to generate realistic blind noise. 
C2N~\cite{hong2020end} adopted a new generator architecture to represent the signal-dependent and spatially correlated noise distribution.
Compared to C2N, our proposed generator has a receptive field limited by the demosaicing method~\cite{zhang2017beyond} and considers quantization noise.

\subsection{Generative Adversarial Network (GAN)}
GAN~\cite{goodfellow2014generative} was proposed to estimate the generative model via simultaneous optimization of the generator and discriminator.
And many researchers have leveraged GANs for image-to-image translation~\cite{isola2017image,dong2017semantic,kaneko2017generative,ledig2017photo,pathak2016context}, whose goal is to translate an input image from one domain to another domain.
However, GAN may suffer from gradient vanishing or exploding during training, and several works~\cite{gulrajani2017improved,lim2017geometric,mao2017least} have proposed proper loss functions to stabilize training.
We build our GAN framework using supervised loss and adversarial loss, thus stabilizing training and achieving promising results.

\subsection{Image Restoration Architecture} 
A popular solution for recovering degraded images is to use CNN-based U-Net-like networks to capture hierarchical information for various image restoration tasks, including image denoising~\cite{yue2020dual,guo2019toward}, deblurring~\cite{kupyn2018deblurgan,kupyn2019deblurgan} and low-light enhancement~\cite{chen2018learning,xia2021deep}. 
Recently, Transformer achieved great advancements in high-level vision problem~\cite{dosovitskiy2020image,liu2021swin,yu2021metaformer} and has also been introduced for image restoration~\cite{chen2021pre,wang2021uformer}.
IPT~\cite{chen2021pre} uses standard Transformer to build a general backbone for various restoration problems. However, IPT is quite huge and requires large-scale datasets. And Uformer~\cite{wang2021uformer} proposed a U-Net-like architecture based on the Swin Transformer~\cite{liu2021swin} for noise and blur removal.
Motivated by MetaFormer~\cite{yu2021metaformer}, we use depth-wise convolution instead of multi-head self-attention to aggregate spatial information. 
The most similar work to ours is NAFNet~\cite{chen2022simple}, which also uses depth-wise convolution and channel attention module to build a MetaFormer-like network.
However, we choose FrozenBN~\cite{wu2021rethinking} instead of LN as the normalization layer.
We also propose a novel channel attention module dubbed ACA, which increases the computational cost slightly compared to the SE modules~\cite{hu2018squeeze} but can aggregate global information more effectively.
\section{Method}
\label{sec:method}
\subsection{Overall Network Architecture}

Our method consists of a new generative network called MPGNet for modeling the UDC imaging degradation process and a U-Net-like network called DWFormer for UDC image restoration.

\begin{figure}[t]
    \centering
    \includegraphics[width=1.0\textwidth]{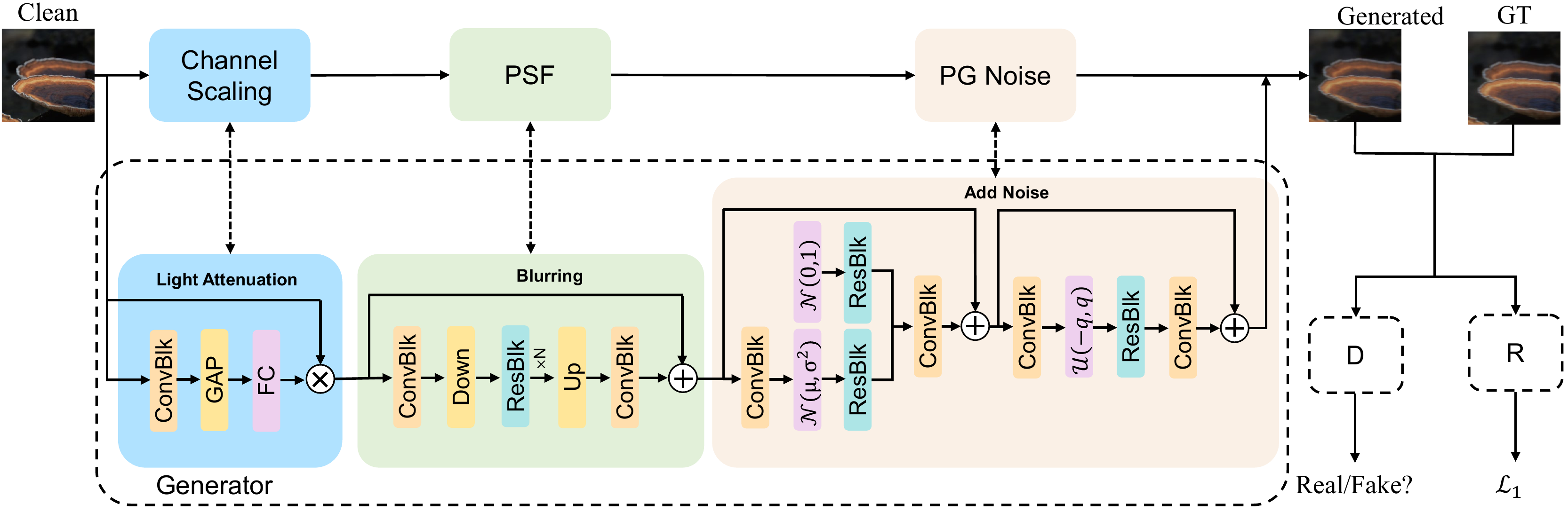}
    \centering
    \caption{Generator network architecture of our proposed MPGNet. We replace each degradation process with a characteristic-related subnetwork to simulate the UDC imaging pipeline and use a GAN-based training scheme for more realistic results.}
    \label{Fig.2}
\end{figure}

\begin{figure}[t]
    \centering
    \includegraphics[width=1.0\textwidth]{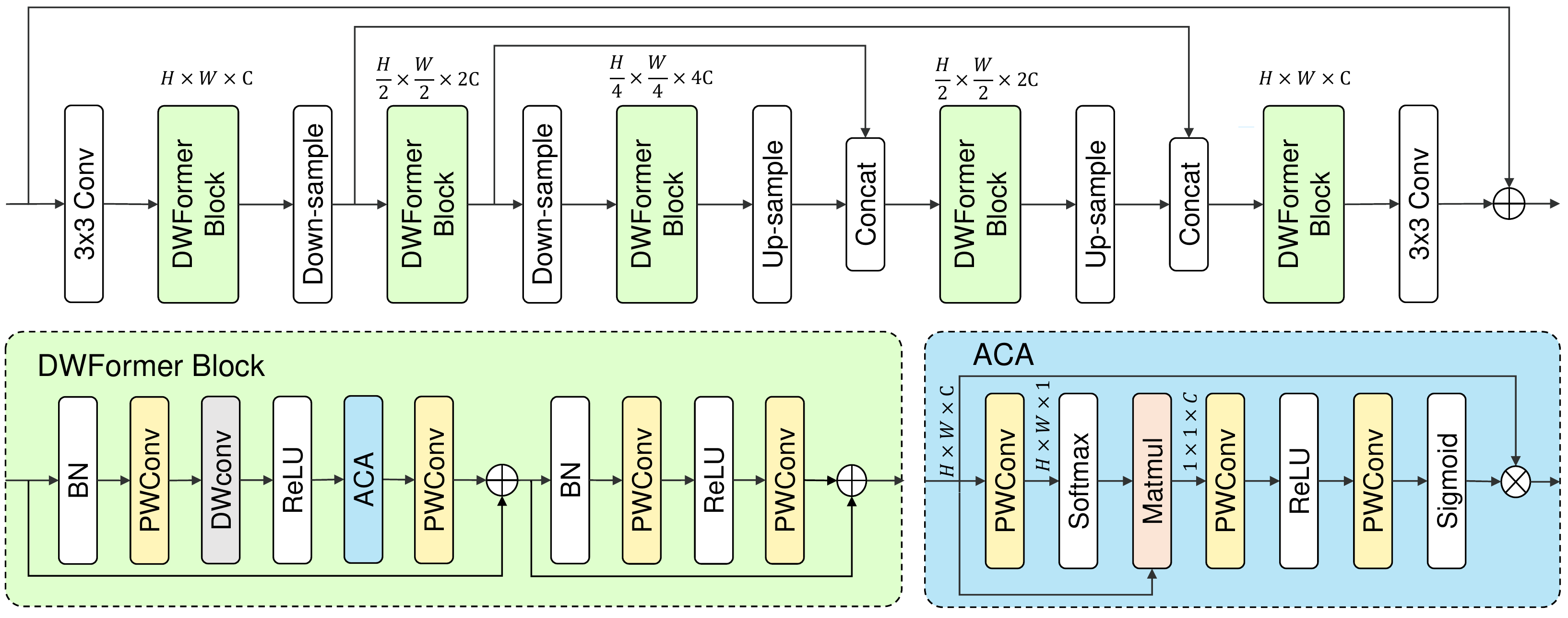}
    \centering
    \caption{Restoration network architecture of our proposed DWFormer.}
    \label{Fig.3}
    \end{figure}

For MPGNet, we adopt GAN framework~\cite{goodfellow2014generative} to improve the realism of generated images. 
As shown in Fig.~\ref{Fig.2}, our generate network architecture consists of a degradation generator and a pixel-wise discriminator. 
The degradation generator MPGNet comprises three parts, \emph{i.e.}, brightness attenuation module, blurring module, and noise module. 
The three parts correspond to channel scaling, diffraction blurring, and Poisson-Gaussian noise corruption in the Statistical Generation Method (SGM)~\cite{zhou2021image}. 
Since the generation process can be considered an image-to-image translation task, we employ a pixel-wise U-Net-like discriminator~\cite{schonfeld2020u} for better performance. 
DWFormer is a U-Net-like network, as shown in Fig.~\ref{Fig.3}, built with our proposed DWFormer Block (DWB).
The DWB evolves from the standard Transformer, and we replace the multi-head self-attention with a depth-wise convolution and channel attention module.

\subsection{MPGNet}

\subsubsection{Brightness Attenuation.}
The brightness attenuation occurs due to organic light-emitting diode (OLED) displays~\cite{li2013four} absorbing part of the light. Since the attenuation is wavelength-related~\cite{nayar1999vision}, the attenuation coefficients to be estimated should be channel-dependent. Besides, brightness is a global statistic and therefore requires global information to be aggregated. To this end, we use several convolution blocks $H_{FE}$ to extract features, a global average pooling $H_{GAP}$ to aggregate information, and a multi-layer perceptron (MLP) to encode attenuation coefficients~\cite{fu2020learning}. We multiply attenuation coefficients of size $1 \times 1 \times C$ with the clean image to obtain the dark image:
\begin{equation}
\begin{aligned}
    F_{BF} &= H_{FE}(x_{clean}),\\
    F_{G} &= H_{GAP}(F_{BF}),\\
    x_{dark} &= x_{clean} \cdot \sigma(H_{D}(\delta(H_{U}(F_{G})))),
\end{aligned}
\end{equation}
where $\delta$ is ReLU and $\sigma$ is Sigmoid function, $H_{D}$ and $H_{U}$ are the channel reduction and channel upsampling operators, respectively~\cite{hu2018squeeze}.

\subsubsection{Blurring.}
The blurring in UDC imaging is caused by light diffraction as the size of the openings in the pixel layout is on the order of the wavelength of visible light.
Although the blur kernel of diffraction could be accurately estimated by computing the Fourier transform of the wavefront function~\cite{voelz2011computational}, it is too complicated, especially when light enters from all directions and forms spatially varying PSFs.
For practical purposes, we choose residual blocks~\cite{he2016deep} as the basis to build the characteristic-related subnetwork to model the blurring process. 
Furthermore, since the blur kernel size is not fixed~\cite{kwon2021controllable}, to better cover the blur kernel of various sizes, we use convolution to downsampling and use sub-pixel convolution~\cite{shi2016real} to upsampling the feature maps.
Finally, we use residual connection to fuse the input and output:
\begin{equation}
    x_{blur} = H_B (x_{dark}) + x_{dark},
\end{equation}
where $H_B$ is the blurring module, $x_{dark}$ is the output of the previous module, and $x_{blur}$ is the blurred and dark image.

\subsubsection{Noise.}
The noise consists of read and shot noise~\cite{hasinoff2014photon}, which are usually formulated as heteroscedastic Gaussian distribution~\cite{kersting2007most}. 
However, noise in the real world is more complicated and spatially correlated, making the heteroscedastic Gaussian distribution model inaccurate. 
Inspired by C2N~\cite{jang2021c2n}, we generate realistic noise by combining signal-independent and signal-dependent noise. 
First, we use a residual block $H_{R_1}$ to transform the noise $n_{s_1} \in \mathbb{R} ^ {h \times w \times d}$ (d=32 is the feature dimension) sampled from the standard normal distribution $\mathcal{N}(0,1)$ to noise $n_i$ with a more complicated distribution:
\begin{equation}
    n_i = H_{R_1}(n_{s_1})~.
\end{equation}
Second, the mean and variance of signal-dependent noise should be highly related to the image signal.
We use a convolutional block to encode the pixel-wise mean $\mu \in \mathbb{R} ^ {h \times w \times d}$ and variance $\sigma \in \mathbb{R} ^ {h \times w \times d}$ from $x_{blur}$.
Since the sampling is not differentiable, we use the reparameterization trick to transform the noise.
Specifically, we sample the noise $n_{s_2} \in \mathbb{R} ^ {h \times w \times d}$ from $\mathcal{N}(0,1)$, and transform it to noise $n_d \thicksim \mathcal{N}(\mu,\sigma ^2)$ via
\begin{equation}
    n_d = H_{R_2}(n_{s_2} \cdot \sigma + \mu),
\end{equation}
where $H_{R_2}$ is also a residual block to transform the distribution of signal-dependent noise to a more complicated distribution.
Considering that the noise is spatially correlated, we use residual blocks with two pixel-wise convolutions and one $3 \times 3$ convolution for both noises. 
After mapping these noises from the initial noise to noise with the target distribution, we take $1 \times 1$ convolution $H_M$ to reduce the dimension to the color space and add them to output $x_{blur}$ of the blurring module: 
\begin{equation}
    x_{noisy} = x_{blur} + H_M (n_i + n_d)~.
\end{equation}
Moreover, recent work~\cite{monakhova2022dancing} shows that the quantification noise significantly impact on low-light imaging. And following the ISP pipeline, the quantization noise should be signal-dependent and added after other noise. Thus we use a convolutional block to encode the pixel-wise quantization noise interval $q \in \mathbb{R} ^ {h \times w \times d}$ from $x_{noisy}$. Also, we use a residual block $H_{R_3}$ to transform the quantization noise $n_{s_3} \in \mathbb{R} ^ {h \times w \times d}$ sampled from the uniform distribution $\mathcal{U}(-q,q)$ to more realistic noise $n_q$:
\begin{equation}
    n_q = H_{R_3}(n_{s_3})~.
\end{equation}
After transforming $n_q$ to the color space by a pixel-wise convolution, we add the quantization noise to the previous noisy image $x_{noisy}$ by a residual connection:
\begin{equation}
    x_{final} = H_{N_q} + x_{noisy},
\end{equation}
where $H_{N_q}$ is the quantization noise module, and $x_{final}$ is the final degraded image.

\subsection{DWFormer}

\subsubsection{Overall Pipeline.}
The overall structure of the proposed DWFormer is a U-Net-like hierarchical network with skip connections. Specifically, given a degraded image $x \in \mathbb{R} ^ {h \times w \times 3}$, DWFormer firstly applied a $3 \times 3$ convolution to extract image features $F_{0} \in \mathbb{R} ^ {h \times w \times c}$. Then, the feature map $F_{0}$ passes through all stages, and each stage contains a stack of proposed DWBs and a downsampling layer or upsampling layer. We use convolution operation to downsampling and sub-pixel convolution to upsampling, respectively. Finally, we also use a $3 \times 3$ convolution to get the residual image $r \in \mathbb{R} ^ {h \times w \times 3}$ and the restored image is obtained by $\hat{x} = y + r$.

\subsubsection{DWFormer Block.}
Transformer~\cite{vaswani2017attention} conducts global self-attention, which leads to quadratic complexity with respect to the number of tokens and has much less inductive bias than CNNs, which may be helpful for low-level tasks. Thus, we modify it to be more suitable for the UDC image restoration task.
Specifically, considering local information is more favorable for noise and blur removal, we use a depth-wise convolution $H_D$ to achieve spatial aggregation and thus significantly reduce the computational cost. 
Further, since brightness correction requires global information for each color channel, we propose an augmented channel attention (ACA) module $H_{ACA}$ to capture global information and place it after the depth-wise convolution. 
Unlike SENet~\cite{hu2018squeeze}, which uses global average pooling to get the global feature descriptor, our proposed ACA starts with a pixel-wise convolution followed by reshape and softmax operations to obtain the weights $W_{p} \in \mathbb{R} ^ {1 \times 1 \times HW}$ of each position of the feature maps $F_{in} \in \mathbb{R} ^ {H \times W \times C}$.
Then we also reshape the $F_{in}$ to $\mathbb{R} ^ {1 \times HW \times C}$ and and use matrix multiplication with $W_{p}$ to obtain the global feature descriptor $F_{C} \in \mathbb{R} ^ {1 \times 1 \times C} $. 
Also, we use a pixel-wise convolution with ReLU activation and a pixel-wise convolution with Sigmoid activation to fully capture channel-wise dependencies. 
Finally, we use the descriptor to rescale the input feature maps to obtain the augmented feature maps.
Besides, we found that BatchNorm (BN) performs better than LayerNorm (LN) in UDC image restoration task when batch size exceeds 16 on a single GPU. Therefore, we choose BN as the normalization layer, and the whole DWFormer block is computed as:
\begin{equation}
\begin{aligned}
    \hat{z}^l &= H_{P_2}(H_{ACA}(\delta(H_D(H_{P_1}(\mathrm{BN}(z^l)))))) + z^l,\\
    z^{l+1} &= \mathrm{MLP}(\mathrm{BN}(\hat{z}^l)) + \hat{z}^l,
\end{aligned}
\end{equation}
where $\delta$ is ReLU and $z^l$ and $z^{l+1}$ denote the input feature maps and output feature maps of DWB respectively. 
Note that we use the FrozenBN~\cite{wu2021rethinking} to avoid the inconsistency in training and testing, \emph{i.e.}, we first train with minibatch statistics in the early training period and use fixed population statistics in the later training period.

\subsection{Training}
Our degradation generator network $G$ is designed to generate a realistic degraded image $\hat{y} \in \mathbb{R} ^ {h \times w \times 3}$ from its clean version $x \in \mathbb{R} ^ {h \times w \times 3}$ and our discriminator network $D$ is designed to distinguish each pixel's real probability. 
Similar to other GAN applications~\cite{isola2017image,wang2018high}, the two networks $G$ and $D$ can be simultaneously optimized in an adversarial way~\cite{goodfellow2014generative} with the least-squares loss function~\cite{mao2017least}:
\begin{equation}
\begin{aligned}
    \mathop{min}\limits_{G}\mathop{max}\limits_{D}\mathcal{L}_{adv}(G,D) & = \mathbb{E}_{y \sim p_{real}(y)}[logD(y)]\\
    & + \mathbb{E}_{x \sim p_{real}(x)}[1-logD(G(x))],
\end{aligned}
\end{equation}
where $y$ is the real degraded image and $x$ is the corresponding clean image.
Here we concat the degraded images with clean images as inputs to $D$. 
And since there have been real paired data~\cite{zhu2017unpaired,isola2017image}, we can also use supervised algorithms to optimize our model.
However, the noise is a distribution-unknown random variable, and if we directly adopted $\mathcal{L} _{1}$ loss on generated and real degraded images, the noise will be eliminated while constructing the image alignment.
To this end, we feed generated degraded images $G(x)$ and real degraded images $y$ into a pre-trained restoration model $R$ to obtain their restored versions and then perform $\mathcal{L}_{1}$ loss between them~\cite{cai2021learning}.
\begin{equation}
    \mathcal{L}_{sup} = || R(G(x)) - R(y) ||_1~.
\end{equation}
We use a hyperparameter $\lambda$ to balance the supervised loss and adversarial loss, and the final loss is:
\begin{equation}
    \mathcal{L} = \mathcal{L}_{adv} + \lambda\mathcal{L}_{sup}~.\label{Loss}
\end{equation}
For convenience, we set the $\lambda$ to 10 for both the T-OLED and P-OLED tracks.
For DWFormer, we just use $\mathcal{L}_{1}$ to train it.

\subsection{Implementation Details}
We train MPGNet for the UDC image generation task and DWFormer for the UDC image restoration task. Thus the training settings are different.
MPGNet is trained with Adam~\cite{kingma2014adam} optimizer ($\beta_1=0.5$, and $\beta_2=0.999$) for $6 \times 10^{4}$ iterations. 
We update the generator once and the discriminator three times in each iteration. 
The initial learning rates are set to $1 \times 10^{-4}$ and $1 \times 10^{-3}$ for generator and discriminator, respectively. 
We employ the cosine annealing strategy~\cite{loshchilov2016sgdr} to steadily decrease the learning rate from an initial value to $1 \times 10^{-6}$ and $1 \times 10^{-5}$ during training. 
The batch size is set to 8, and we randomly perform horizontal and vertical flips for data augmentation. 
And DWFormer has five stages, and the number of DWB in each layer is $\{8, 8, 8, 6, 6\}$, respectively. 
The model is also trained with Adam optimize ($\beta_1=0.9$, and $\beta_2=0.999$) for $2 \times 10^{5}$ iterations. 
The initial learning rate is set to $1 \times 10^{-4}$ and the cosine annealing strategy is adopted to steadily decrease the learning rate to $1 \times 10^{-6}$. 
The batch size is 64, and we achieve data augmentation by randomly performing flips and rotations.

\section{Experiment}
\subsection{Dataset}
The real image pairs used in the experiments are the P-OLED and T-OLED datasets~\cite{zhou2021image} provided by UDC 2020 Image Restoration Challenge.
Both datasets have 300 clean-degraded image pairs of size $1024 \times 2048$.
Also, we leverage the high-resolution DIV2K dataset~\cite{agustsson2017ntire} captured under real environments to generate realistic degraded images. 
The DIV2K dataset provides 900 clean images. Similar to prior work~\cite{zhou2021image}, we optionally rotate and resize the clean images and generate clean-degraded image pairs with the resolution of $1024 \times 2048$. 
The real and generated image pairs are cropped into patches of size $256 \times 256$ and are randomly sampled to gather training mini-batches. 

\begin{figure}[t]
\centering
\includegraphics[width=1.0\textwidth]{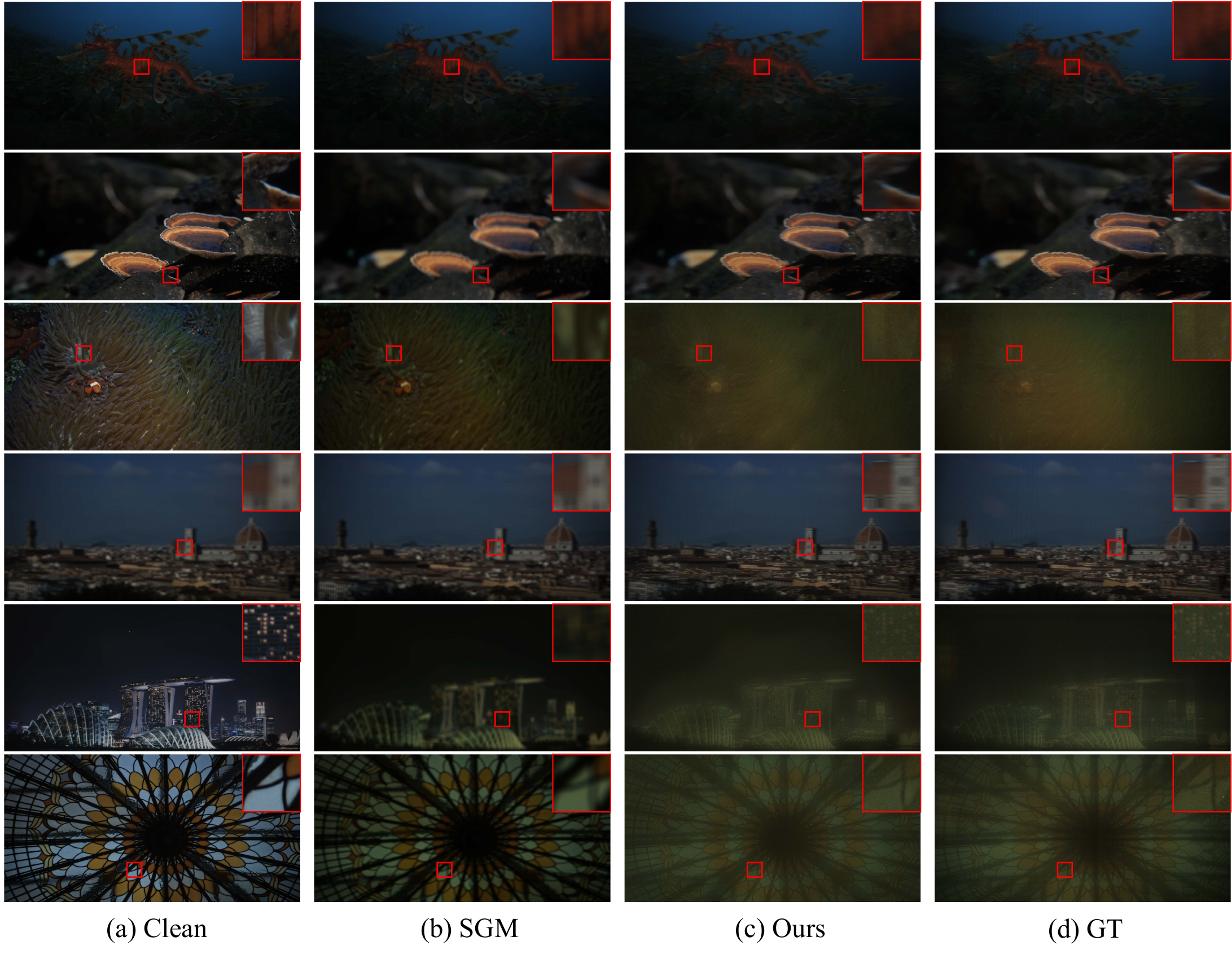}
\caption{
Visual comparison of the generated degraded image samples on the UDC dataset. (a) A clean image. (b-e) Generated degraded image of SGM and Ours, respectively. (f) The ground truth degraded image. The top half of the figure is on the T-OLED track, and the bottom half of the figure is on the P-OLED track. We amplify the images on the P-OLED track for comparison.}
\label{Fig.4}
\end{figure}

\begin{figure}[t]
\centering
\includegraphics[width=1.0\textwidth]{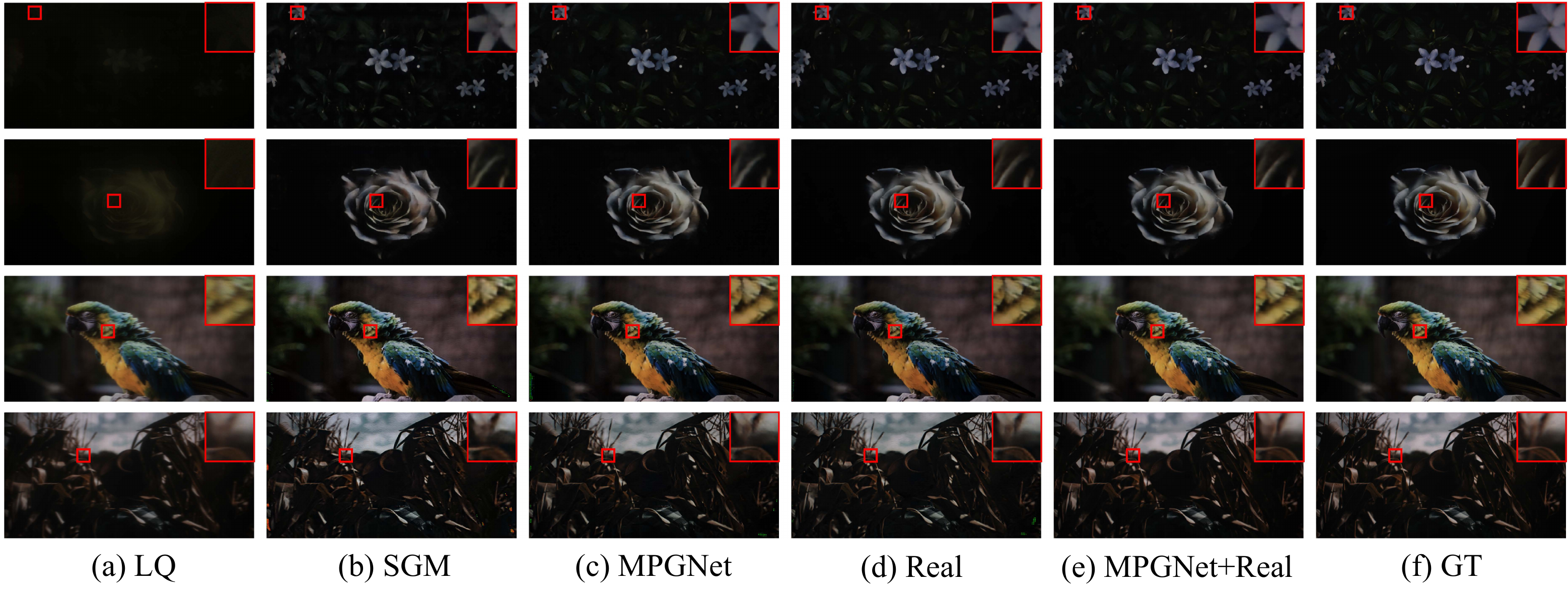}
\caption{Comparison between restoration results on the UDC dataset. (a) The degraded image. (b-c) The clean images are restored using the models trained by the different datasets. (f) The ground-truth. The top two are from the T-OLED track, and the bottom two are from the P-OLED track.}
\label{Fig.5}
\end{figure}

\subsection{Comparison of Generators}
To holistically evaluate the quality of MPGNet-generated images, we employ several tactics.
First, we present several qualitative examples for perceptual studies in Fig.~\ref{Fig.4}.
The results demonstrate that the SGM-generated~\cite{zhou2021image} degraded images cannot accurately estimate the blur and noise distribution. In contrast, the MPGNet-generated results are closer to the ground truths and preserve the degradation's diversity.

Second, we use different generated datasets to train the restoration models and evaluate them on the UDC benchmark.
Fig.~\ref{Fig.5} shows that the model trained with the SGM-generated dataset yields still blurry and brightness-unaligned results, while the model trained with the MPGNet-generated dataset produces results closer to the ground truth.
Also, Table~\ref{tab1} illustrates that our method outperforms SGM by 3.31 dB on the P-OLED track and 2.49 dB on the T-OLED track by only using the synthetic dataset. We can further improve the restoration model's performance by using both generated and real datasets for training, implying that our generated data can complement the real data and thus enhance the model's generalization. 

Intuitively, using a single model as a generator is more common. 
Thus we replace the entire generation model with an end-to-end U-Net~\cite{agustsson2017ntire}, which has a competitive number of parameters and computation cost to MPGNet. 
However, the U-Net performs poorly. We believe this is mainly because a single network tends to confuse multiple degradation processes, leading to model convergence to a poor local optimum. 
We assign the degradation process to multiple characteristic-related sub-networks, which dramatically avoids this local optimum, showing that physical constraints work.
Also, from the results, we find that SGM performs worse than U-Net, which may be due to the inaccurate parameters from the manual statistics. We argue that our MPGNet provides a solution to avoid it.

\setlength{\tabcolsep}{8pt}{
\begin{table}[t]
\caption{Performance comparison between other methods and our MPGNet.}
\begin{center}
\renewcommand\arraystretch{1.25}
\begin{tabular}{lllll}
\hline
\multirow{2}*{{Methods}}&\multicolumn{2}{c}{{P-OLED}}&\multicolumn{2}{c}{{ T-OLED}}\\
\cline{2-5}
&\hspace{6pt}{PSNR} & \hspace{10pt}{SSIM} & \hspace{6pt}{PSNR} & \hspace{10pt}{SSIM}\\
\hline
{Real}&{33.21} & {0.960}&{38.96} &{0.984}\\
{SGM}&{28.61}$_{\downarrow 4.60}$ & {0.911}$_{\downarrow 0.049}$&{34.82}$_{\downarrow 4.14}$ & {0.950}$_{\downarrow 0.034}$\\
{SGM + Real}&{32.56}$_{\downarrow 0.65}$ & {0.957}$_{\downarrow 0.003}$&{38.42}$_{\downarrow 0.54}$ & {0.980}$_{\downarrow 0.004}$\\
{U-Net}&{30.41}$_{\downarrow 2.80}$ & {0.912}$_{\downarrow 0.048}$&{36.20}$_{\downarrow 2.76}$ & {0.958}$_{\downarrow 0.026}$\\
{MPGNet}&{31.95}$_{\downarrow 1.26}$ & {0.941}$_{\downarrow 0.019}$&{37.33}$_{\downarrow 1.63}$ & {0.970}$_{\downarrow 0.014}$\\
{MPGNet + Real}&{34.22}$_{\uparrow 1.01}$ & {0.964}$_{\uparrow 0.004}$&{39.55}$_{\uparrow 0.59}$ &{0.986}$_{\uparrow 0.002}$\\
\hline
\end{tabular}
\label{tab1}
\end{center}
\end{table}
}

\subsection{Comparison of Restoration Models}
We compare the performance of the proposed DWFormer with other learning-based approaches on the UDC benchmark, and the quantitative comparisons in terms of PSNR and SSIM metrics are summarized in Table~\ref{tab2}. 
The results show that our method achieves state-of-the-art results for both the P-OLED track and T-OLED track. 
In particular, our DWFormer achieves 0.22 and 0.12 dB PSNR improvements over the previous best methods PDCRN~\cite{panikkasseril2020transform} and DRANet~\cite{nie2020dual} on the P-OLED and T-OLED tracks, respectively. 
Using both generated and real data to train our model, DWFormer can improve 1.01 dB on the P-OLED track and 0.59 dB on the T-OLED track over the previous one.

\setlength{\tabcolsep}{8pt}{
\begin{table}[t]
\caption{
Restore qualitative evaluation on the UDC benchmark. Note that $^\dagger$ means that the models are trained with both generated and real datasets.}
\begin{center}
\renewcommand\arraystretch{1.25}
\begin{tabular}{lcccccc}
\hline
\multirow{2}*{{Methods}}&\multicolumn{2}{c}{{P-OLED}}&\multicolumn{2}{c}{{ T-OLED}}&\multicolumn{2}{c}{{Overhead}}\\
\cline{2-7}
&{PSNR} & {SSIM} & {PSNR} & {SSIM}&{Params}&{MACs}\\
\hline
{ResNet~\cite{zhou2021image}}&{27.42} & {0.918}&{36.26} & {0.970}&\textcolor{blue}{{1.37M}}&{40.71G}\\
{UNet~\cite{zhou2021image}}&{29.45} & {0.934}&{36.71} & {0.971}&{8.94M}&{17.09G}\\
{SGDAN~\cite{zhou2020udc}}&{30.89} & {0.947}&{36.81} & {0.971}&{21.1M}&{\textcolor{blue}{7.25G}}\\
{ResUNet~\cite{yang2020residual}}&{30.62} & {0.945}&{37.95} & {0.979}&{16.50M}&{15.79G}\\
{DRANet~\cite{nie2020dual}}&{31.86} & {0.949}&{\textcolor{blue}{38.84}} & \textcolor{blue}{{0.983}}&{79.01M}&{168.98G}\\
{PDCRN~\cite{panikkasseril2020transform}}&\textcolor{blue}{{32.99}} & \textcolor{blue}{{0.958}}&{37.83} & {0.978}&{3.65M}&\textcolor{red}{{6.31G}}\\
{RDUNet~\cite{yang2020residual}}&{30.12} & {0.941}&{38.22} & {0.980}&{47.93M}&{46.01G}\\
{DWFormer(Ours)}&\textcolor{red}{{33.21}} & \textcolor{red}{{0.960}}&\textcolor{red}{{38.96}} &{\textcolor{red}{0.984}}&\textcolor{red}{{1.21M}}&{13.46G}\\
\hline
{PDCRN$^\dagger$}&{\textcolor{blue}{34.02}} & {\textcolor{blue}{0.962}}&{38.75} & {0.982}&\textcolor{blue}{{3.65M}}&\textcolor{red}{{6.31G}}\\
{RDUNet$^\dagger$}&{32.45} & {0.952}&{\textcolor{blue}{39.05}} & \textcolor{blue}{{0.984}}&{47.93M}&{46.01G}\\
{DWFormer$^\dagger$(Ours)}&{\textcolor{red}{34.22}} & {\textcolor{red}{0.964}}&{\textcolor{red}{39.55}} & \textcolor{red}{{0.986}}&{\textcolor{red}{1.21M}}&\textcolor{blue}{{13.46G}}\\
\hline
\end{tabular}
\label{tab2}
\end{center}
\end{table}
}

\subsection{Ablation Studies}

\setlength{\tabcolsep}{6pt}
{
\begin{table}[t]
\caption{Different modules' effects on the MPGNet's performance. The performance is evaluated on the generated datasets using our DWFormer.}
\begin{center}
\renewcommand\arraystretch{1.25}
\begin{tabular}{lllll}
\hline
\multirow{2}*{{Methods}}&\multicolumn{2}{c}{{ P-OLED}}&\multicolumn{2}{c}{{ T-OLED}}\\
\cline{2-5}
&\hspace{6pt}{PSNR} & \hspace{7pt} {SSIM} & \hspace{6pt} {PSNR} & \hspace{7pt} {SSIM}\\
\hline
{MPGNet}& 31.95 &0.941&37.33 & 0.970\\
{MPGNet w/ SGM-Light}&$30.81_{\downarrow 1.14}$ & {0.930}$_{\downarrow 0.011}$ &{36.96}$_{\downarrow 0.37}$ & {0.965}$_{\downarrow 0.005}$\\
{MPGNet w/ SGM-Blur}&{30.10}$_{\downarrow 1.85}$ & {0.915}$_{\downarrow 0.026}$&{35.87}$_{\downarrow 1.46}$ & {0.959}$_{\downarrow 0.011}$\\
{MPGNet w/ SGM-Noise}&{31.16}$_{\downarrow 0.79}$ & {0.934}$_{\downarrow 0.007}$&{36.30}$_{\downarrow 1.03}$ & {0.962}$_{\downarrow 0.008}$\\
{MPGNet w/o $n_q$}&{31.36}$_{\downarrow 0.59}$ & {0.935}$_{\downarrow 0.006}$&{36.87}$_{\downarrow 0.46}$ & {0.964}$_{\downarrow 0.006}$\\
{MPGNet w/o $n_d$}&{31.26}$_{\downarrow 0.69}$ & {0.935}$_{\downarrow 0.006}$&{36.55}$_{\downarrow 0.78}$ &{0.963}$_{\downarrow 0.007}$\\
{MPGNet w/o $n_i$}&{31.64}$_{\downarrow 0.31}$ & {0.937}$_{\downarrow 0.004}$&{37.04}$_{\downarrow 0.29}$ & {0.967}$_{\downarrow 0.003}$\\
\hline
\end{tabular}
\label{tab3}
\end{center}
\end{table}
}

\setlength{\tabcolsep}{6pt}
{
\begin{table}[t]
\caption{Different modules' effects on the DWFormer's performance.}
\begin{center}
\renewcommand\arraystretch{1.25}
\begin{tabular}{llllll}
\hline
\multirow{2}*{{Methods}}&\multicolumn{2}{c}{{ P-OLED}}&\multicolumn{2}{c}{{ T-OLED}}&\multirow{2}*{{MACs}}\\
\cline{2-5}
&\hspace{6pt}{PSNR} & \hspace{7pt} {SSIM} & \hspace{6pt} {PSNR} & \hspace{7pt} {SSIM}\\
\hline
{DWFormer}& 33.21 &0.960&38.96 & 0.984 & 13.46G\\
{ACA \hspace{0pt} $\rightarrow$ None}&$32.62_{\downarrow 0.59}$ & {0.951}$_{\downarrow 0.009}$ &{38.45}$_{\downarrow 0.51}$ & {0.980}$_{\downarrow 0.004}$& 13.42G\\
{ACA \hspace{0pt} $\rightarrow$ SE}&$33.00_{\downarrow 0.21}$ & {0.958}$_{\downarrow 0.002}$ &{38.72}$_{\downarrow 0.24}$ & {0.982}$_{\downarrow 0.002}$& 13.43G\\
{DWB $\rightarrow$ Swin}&$33.07_{\downarrow 0.14}$ & {0.959}$_{\downarrow 0.001}$ &{38.84}$_{\downarrow 0.12}$ & {0.982}$_{\downarrow 0.002}$& 16.22G\\
{BN \hspace{6pt} $\rightarrow$ LN}&$33.11_{\downarrow 0.10}$ & {0.957}$_{\downarrow 0.003}$ &{38.90}$_{\downarrow 0.06}$ & {0.982}$_{\downarrow 0.002}$&13.46G\\

\hline
\end{tabular}
\label{tab5}
\end{center}
\end{table}
}

We first study the impact of each module in MPGNet on performance.
Specifically, we replace brightness correction, blurring, and noise modules with corresponding modules in SGM.
Further, we remove each noise component, \emph{i.e.}, quantization noise $n_q$, signal-dependent noise $n_d$ and signal-independent noise $n_i$ to explore their effects.
We use these modified networks to generate datasets that are used to train restoration models. 
And we evaluate the restoration models on the UDC benchmark as shown in Table~\ref{tab3}.

As we can see, the blurring module has the most significant impact on the model performance, indicating that diffraction caused by hardware structure is the most critical degradation factor in the UDC imaging process. 
And ablation studies of noise and brightness modules show that different materials' main degradation components differ.
Surprisingly, quantization noise plays such a significant role in the noise module. 
It may be due to the brightness attenuation during the imaging process and the low number of image bits. 
Also, such a phenomenon indicates that the actual noise module of the UDC images is quite complex and challenging to simulate using only Poisson-Gaussian distribution.

To verify the effectiveness of the DWFormer modules, we performed ablation studies for the normalization, depth-wise convolution, and attention mechanism module, and Table~\ref{tab3} shows the results.
We notice that the removal of ACA causes a significant performance drop, implying that global information is indispensable. 
And our ACA is much better than SE with negligible additional computation cost.
Also, the performance decreased if we used Swin Transformer's block, indicating that the locality properties provided by depth-wise convolution work for the UDC restoration task.
Besides, BatchNorm surpasses LayerNorm.
It is worth noting that BatchNorm can be fused into a convolutional layer when inferring, which makes the network run faster than the network with LayerNorm.

\begin{figure}[t]
    \centering
    \includegraphics[width=1.0\textwidth]{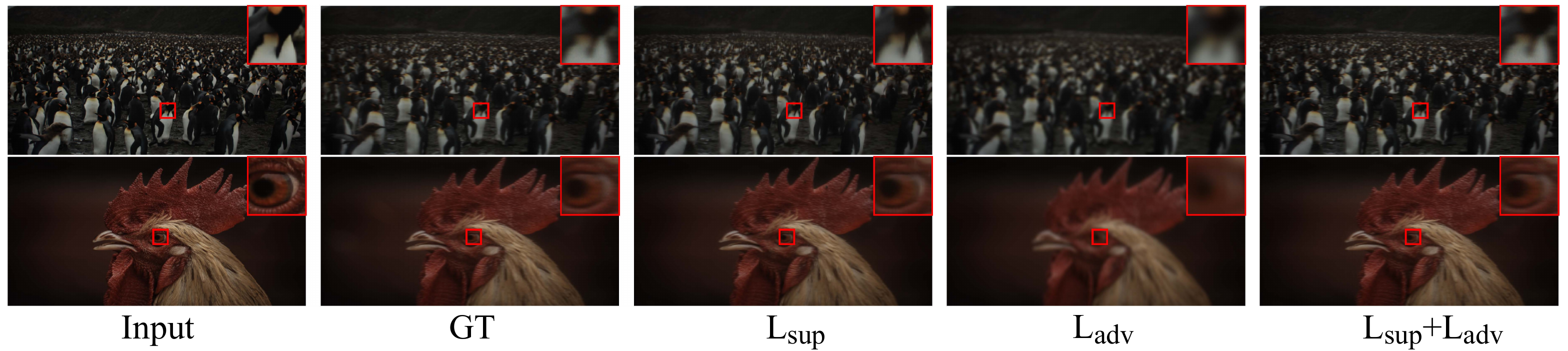}
    \caption{Different losses lead to different generation results. Each column shows the results of training at different losses.}
    \label{Fig.6}
\end{figure}

Moreover, we explore the effect of $\mathcal{L}_{sup}$ and $\mathcal{L}_{adv}$ and use them for training MPGNet. Fig.~\ref{Fig.6} shows the generated results.
$\mathcal{L}_{sup}$ alone leads to reasonable but noiseless results. 
$\mathcal{L}_{adv}$ alone gives much blurrier results. 
It is because the $\mathcal{L}_{sup}$ loss will still eliminate the noise while constructing the image alignment though we have attempted to alleviate it, and the weak constraint of GAN causes the content restoration to be more difficult. 

\setlength{\tabcolsep}{12pt}{
\begin{table}[t]
\caption{Different loss terms for training MPGNets. The performance is evaluated on the generated datasets using our DWFormer.}
\begin{center}
\renewcommand\arraystretch{1.25}
\begin{tabular}{lcccc}
\hline
\multirow{2}*{{Methods}}&\multicolumn{2}{c}{{ P-OLED}}&\multicolumn{2}{c}{{ T-OLED}}\\
\cline{2-5}
&{PSNR} & {SSIM} & {PSNR} & {SSIM}\\
\hline
{$\mathcal{L}_1$}&{31.13} & {0.908}&{35.98} & {0.951}\\
{$\mathcal{L}_{sup}$}&{31.23} & {0.911}&{36.08} & {0.966}\\
{$\mathcal{L}_{adv}$}&{31.55} & {0.945}&{36.02} & {0.957}\\
{$\mathcal{L}_{adv}$ + $\mathcal{L}_{sup}$}&\textcolor{red}{{33.22}} & \textcolor{red}{{0.960}}&{38.80} &{0.981}\\
{$\mathcal{L}_{adv}$ + 10$\mathcal{L}_{sup}$}&\textcolor{blue}{{33.21}} & \textcolor{blue}{{0.960}}&\textcolor{blue}{{38.96}} & \textcolor{blue}{{0.984}}\\
{$\mathcal{L}_{adv}$ + 100$\mathcal{L}_{sup}$}&{33.03} & {0.956}&{\textcolor{red}{38.98}} & \textcolor{red}{{0.984}}\\
\hline
\end{tabular}
\label{tab4}
\end{center}
\end{table}
}

We further use these generated datasets to train our DWFormer and evaluate them on the UDC benchmark for quantitative assessment, and the results are shown in Table~\ref{tab4}.
Note that $\mathcal{L}_1$ means that we directly compute the $\mathcal{L}_1$ loss on the generated image. 
In contrast, $\mathcal{L}_{sup}$ means we take the generated images through a restoration network and then calculate the loss. 
The $\mathcal{L}_1$ is worse than the $\mathcal{L}_{sup}$, implying that the latter can preserve the random noise to some extent.
Note that if we do not feed clean images as extra inputs to $D$, MPGNet fails to generate valid degraded images and falls into a mode collapse. 

Here, we use both $\mathcal{L}_{sup}$ and $\mathcal{L}_{adv}$ to boost the model performance. 
And We further evaluate it with different ratios (changing the value of $\lambda$). 
It is found that the optimal ratio is highly correlated with the data itself, and different datasets hold different optimal ratios. 
Experiments show that $\mathcal{L}_{adv}$ plays a more significant role on the P-OLED track because the dataset is severely degraded.
And the degradation is diminished on the T-OLED track, so the weight of $\mathcal {L}_1$ needs to be increased to ensure content consistency.

\section{CONCLUSION}
In this paper, we start from the degradation pipeline of the UDC imaging and replace each degradation process with a subnetwork, which forms our MPGNet. 
Further, the GAN framework is adopted to generate more realistic degraded images. 
Based on the analysis of UDC image degradation, we propose a novel block modified from Transformer and use it to build a U-Net-like image restoration network named DWFormer. 
Experiments show that our MPGNet generates more realistic degraded images than previous work, and DWFormer achieves superior performance. 
Finally, We use both generated and real datasets to train DWFormer, and further boost its performance, showing our generated data can be complementary to real data.
\bibliographystyle{splncs}
\bibliography{egbib}

\end{document}